\documentclass[submission,copyright,creativecommons, colorlinks=true, linkcolor=Purple, citecolor=Purple, urlcolor=Purple ]{eptcs}
\providecommand{\event}{ICLP 2020} 
\usepackage{mathptmx}

\title
{Justifications for Goal-Directed\\
  Constraint Answer Set Programming\thanks{Work partially supported by
    MINECO project TIN2015-67522-C3-1-R (TRACES), MICINN project
    PID2019-108528RB-C21 (ProCode), Comunidad de Madrid project
    S2018/TCS-4339 BLOQUES-CM co-funded by EIE Funds of the European
    Union, NSF awards IIS 1718945, IIS 1910131, IIP 1916206, and a
    DARPA award.}}  \def\titlerunning{Justifications for s(CASP)}

\author{%
Joaqu\'in Arias$^{1}$ \qquad\qquad Manuel Carro$^{1,2}$
\institute{$^1$IMDEA Software Institute \\
  $^2$Universidad Polit\'ecnica de Madrid}
\email{joaquin.arias@imdea.org}\\[-1.5em]
\email{manuel.carro@\{imdea.org,upm.es\}}
  \and
Zhuo Chen$^3$ \qquad\qquad Gopal Gupta$^3$
\institute{$^{3}$University of Texas at Dallas\\~}
\email{\{zhuo.chen,gupta\}@utdallas.edu}
}
\def\authorrunning{Joaqu\'{\i}n Arias et al.}

\usepackage[T1]{fontenc}

\usepackage{amssymb}
\usepackage{amsmath}
\usepackage{textcomp}
\usepackage{calc}
\usepackage{comment}
\usepackage{enumitem}
\usepackage[justification=centering]{caption}
\usepackage[skip=5pt,margin=5pt]{subcaption}
\usepackage{mathptmx} 
\usepackage{hyperref}

\usepackage{url}
\usepackage{relsize}
\usepackage{multicol}
\usepackage{multirow}
\usepackage{paracol}
\usepackage{mdframed}
\usepackage{longtable}
\usepackage{booktabs}
\usepackage{graphicx} 
\usepackage[dvipsnames]{xcolor}
\definecolor{PrologPredicate}{RGB}{0,0,200}
\definecolor{PrologVar}      {RGB}{145,032,039}
\definecolor{PrologComment}  {RGB}{169,082,044}
\definecolor{PrologOther}    {rgb}{0.2,0.2,0.2}
\definecolor{PrologString}   {RGB}{070,120,200}
\usepackage{listings} 
\usepackage{wrapfig}
\newcommand{\code}{\lstinline[style=MyInline]}
\lstdefinestyle{MyInline}
{
  basicstyle = \ttfamily\color{PrologOther},
  breaklines = true,
  breakatwhitespace=true,
  upquote = true,
  literate =
  {,}{}{0\discretionary{,}{}{,}}
  {\ │}{{$\mid$}}1 
  {|}{{$\mid$}}1
  {\\\{}{{\{}}1
  {\\\}}{{\}}}1
  {[}{{\small[}}1
  {]}{{\small]}}1
  {.=.}{{\#=}}3
  {.<.}{{\#<}}3
  {.>.}{{\#>}}3
  {.=<.}{{\#=<}}4
  {.>=.}{{\#>=}}4
  {<}{{<}}2
  {>}{{>}}2
  {=<}{{=<}}3
  {>=}{{>=}}3
  {\ \\=}{{\,\char"5C=\,}}2
  {\\=}{{\char"5C=}}2
  {?-}{{?-\,}}3
  {:-}{{:-\,}}3
  {\\$}{{\$}}1
}
\lstdefinestyle{Plain}
{
  keywords = {},
  upquote = true,
  basicstyle = \relsize{-0.5}\ttfamily\color{PrologPredicate},
  basewidth = 0.48em,
  numbers=none,
  xleftmargin=0cm,
  moredelim = {*[s][\color{black!40!PrologPredicate}]{\#pred}{.}},
  moredelim = {*[s][\color{black!40!PrologPredicate}]{\#show}{.}},
  moredelim = {*[s][\color{black!40!PrologPredicate}]{\#hide}{.}},
  moredelim = {*[s][\color{PrologVar}]{(}{)}},
  moredelim = {*[s][\color{PrologString}]{'}{'}},
 commentstyle = \mdseries\color{PrologComment},
  morecomment=[l]\%,
   literate     =
  {|}{{$\mid$}}1
  {\\$}{{\$}}1
  {\ │}{{$\mid$}}1,
}
\lstdefinestyle{MySCASP}
{
  keywords = {},
  upquote = true,
  basicstyle = \relsize{-0.5}\ttfamily\color{PrologPredicate},
  basewidth = 0.48em,
  moredelim = {**[is][\color{PrologComment}]{`}{`}},
  moredelim = {*[s][\color{black!40!PrologPredicate}]{\#pred}{.}},
  moredelim = {*[s][\color{black!40!PrologPredicate}]{\#show}{.}},
  moredelim = {*[s][\color{black!40!PrologPredicate}]{\#hide}{.}},
  moredelim = {*[s][\color{PrologVar}]{(}{)}},
  moredelim = {*[s][\color{PrologString}]{'}{'}},
  moredelim = {*[s][\color{PrologOther}]{:-}{.}},
  moredelim = {*[s][\color{red}]{/*}{*/}},
  commentstyle = \mdseries\color{PrologComment},
  morecomment=[l]\%,
  morecomment=[s]{/*}{*/},
  literate     =
  {|}{{$\mid$}}1
  {\ │}{{$\mid$}}1
  {[}{{\color{PrologOther}\small[}}1
  {]}{{\color{PrologOther}\small]}}1
  {\\$}{{\$}}1
  {&(}{{\color{PrologOther}(}}1
  {&)}{{\color{PrologOther})}}1
  {&.}{{.}}0
  {\\=}{{\char"5C=}}2
  {\\$}{{\$}}1,
}
\usepackage{xspace}
\usepackage[
disable,
textwidth=0.2\textwidth,textsize=footnotesize]{todonotes}
\newcommand{\jahcommin}[1]{\todo[inline,backgroundcolor=green,linecolor=yellow]{\textsf{JAH:
      #1}}\xspace}
\newcommand{\mclcomm}[1]{\todo[bordercolor=yellow,linecolor=yellow]{\textsf{MCL:
      #1}}\xspace}
\newcommand{\mclcommin}[1]{\todo[inline,bordercolor=yellow,linecolor=yellow]{\textsf{MCL: #1}}}

\newtheorem{definition}{Definition}

\newtheorem{example}{Example}

\newcommand{\redaftfig}{\vspace*{0em}}
\newcommand{\redafttab}{\vspace*{0em}}
\newcommand{\redbefsect}{\vspace{0em}}
\newcommand{\redbefmult}{\vspace*{-.75em}}
\newcommand{\redaftmult}{\vspace*{-1em}}
\newcommand{\mycite}[1]{}
\newcommand{\myurl}[1]{\href{http://www.cliplab.org/papers/sCASP-ICLP2020/#1}{\nolinkurl{#1}}}

\newcommand\lrul[2]{\ensuremath{#1\ \leftarrow \ #2}}

\begin{document}

\label{firstpage}

\maketitle

\vspace{-1em}
\begin{abstract}
  \hyphenpenalty 10000 %
  \exhyphenpenalty 10000 %

  Ethical
  and legal
  concerns make it necessary for programs that may directly influence
  the life of people (via, e.g., legal or health counseling)
  to \emph{justify} in human-understandable terms the advice given.
  %
  %
  Answer Set Programming has a rich semantics that makes it possible
  to very concisely express complex knowledge.
  %
  However, justifying why an answer is a consequence from an ASP
  program may be non-trivial --- even more so when the user is an
  expert in a given domain, but not necessarily knowledgeable in
  ASP. 
  %
  %
  Most ASP systems generate answers using SAT-solving procedures on
  ground rules that do not match how humans perceive reasoning.
  %
  %
  We propose using s(CASP), a query-driven, top-down execution model
  for predicate ASP with constraints to generate justification trees
  of (constrained) answer sets.  The operational semantics of s(CASP)
  relies on backward chaining, which is intuitive to follow and lends
  itself to generating explanations that are easier to translate into
  natural language.
  We show how s(CASP) provides minimal justifications for, among
  others, relevant examples proposed in the literature, both as search
  trees but, more importantly, as explanations in natural language.
  We validate our design with real ASP applications and evaluate the
  cost of generating s(CASP) justification trees.
\end{abstract}


\redbefsect
\section{Introduction and Motivation}
\label{sec:intr-motiv}

Since the European Union approved the General Data Protection
Regulation (GDPR)~\cite{GDPR}, every person affected by a decision
made by an automated process has the right to obtain an explanation
of the decision reached and to challenge the decision (Art.~71).
Therefore, the uptake of current Artificial Intelligence (AI)
systems is restricted by their limitation to explain 
their decisions
in a way amenable to human understanding.
For example, AI systems based on machine learning 
may give accurate outcomes, but they work as a black box without providing an
intuitive high-level description of how they reach a decision.  It is
therefore hard for users and/or programmers to understand or verify
the principles that govern them.
In the context of Explainable Artificial
Intelligence~\cite{2017explainable} (XAI), the USA Department of Defense
(DoD) is facing challenges that demand XAI to understand,
appropriately trust, and effectively manage an emerging generation of
artificially intelligent machine partners.

Answer Set Programming (ASP) is a successful paradigm for developing
intelligent applications and has attracted much attention due to its
expressiveness, ability to represent knowledge, incorporate
non-monotonicity, and model combinatorial problems.
ASP uses the stable model semantics~\cite{gelfond88:stable_models} for
programs with negation.
It is a declarative paradigm where the programmer specifies rules that
describe the problem to be solved and the ASP system computes the
solution. The solution of the program is an answer set that satisfies
the rules of the program under the stable model semantics.
Most ASP systems follow bottom-up executions that require a grounding
phase where the variables of the program are replaced with their
possible values. During the grounding phase, links between variables
are lost and therefore an explanation framework for these systems
must face many challenges to provide a concise justification of why a
specific answer set satisfies the rules (and which rules). Some of the
most relevant approaches to justification are:

\begin{itemize} 
\item Off-line and on-line
  justifications~\cite{DBLP:journals/tplp/PontelliSE09} provide a
  graph-based explanation of the truth value (i.e., true, false, or
  assume) of a literal.
  The explanation \emph{assume} is used for literals whose truth value
  is not being requested.  This proposal provides a basic theory to
  support debugging of ASP programs.
\item Causal Graph Justification~\cite{DBLP:journals/tplp/CabalarFF14}
  explains why a literal is contained in an answer set, but not why a
  negated literal is not contained. It can be used to formalize and
  reason with causal knowledge (i.e., it relies on a causal
  interpretation of rules and the idea of causal chain).
\item Labeled ABA-Based Answer Set Justification,
  LABAS~\cite{DBLP:journals/tplp/0001T16} explains the truth value of
  an extended literal with respect to a given answer set. A literal is
  in the answer set if a derivation of this literal is supported and
  is not in the answer set if all derivations of this literal are
  ``attacked''. I.e., a literal is explained in terms of the (negated)
  literals necessary for its derivations, rather than in terms of the
  whole answer set.
\end{itemize}

\noindent
However, these approaches are applied to grounded versions of the
programs and they may produce unwieldy justifications.  The
explanation of answers from ASP programs with constraint (both under
discrete or dense domains) is even more challenging since the
grounding phase removes the relationships among constrained variables,
in the case of dense domains, because of the difficulty of
representing the ranges of the variables.

On the other hand, systems that follow a top-down execution  can 
trace which rules have been used to obtain the answers more easily.
One such system is 
ErgoAI (\url{https://coherentknowledge.com}), based on
XSB~\cite{xsb-journal-2012}, that generates justification trees
for programs with variables. ErgoAI has been applied to analyze
streams of financial transactions in near real-time providing
explanations in English that are fully detailed and interactively
navigable. However, default negation in ErgoAI is based on the
well-founded semantics~\cite{VanGelder91}
and therefore ErgoAI 
is not a framework that can explain the answers from ASP
programs. \mclcomm{Note to self: check the real difference.  Is WFS a
  particular type of stable semantics?}


In this work we propose the use of s(CASP)~\cite{scasp-iclp2018}, a
goal-directed interpreter for ASP with constraints (CASP), to explain
the results of CASP programs.
%
%
The top-down evaluation does not require a grounded program.
Moreover, the execution of an s(CASP) program returns partial stable
models that are the relevant subsets of the ASP stable models which
include only the (negated) literals necessary to support the initial
query.  To the best of our knowledge, s(CASP) is the only system that
exhibits the property of relevance~\cite{pereira89:relevant-counterfactuals}.

This paper shows how s(CASP) can use top-down evaluation trees to
generate minimal justifications where it is possible to control which
literals should appear.  We also provide a mechanism to present the
justifications with natural language using a generic translation and
the possibility of customizing it with directives that provide
translation patterns.  Both plain text and user-friendly, expandable
HTML can be generated.
%
%
Additionally, the translation into natural language can be used with
the program text itself, thereby making it easier for experts without
a programming background to understand both the program and the results
of its execution.


%
We validate the design and the expressiveness of the s(CASP)
justification trees with examples from the literature and with complex
applications such as 
a Physician Advisory System for Chronic Heart Failure Management.
Finally, we evaluate the cost of generating the s(CASP) justification
trees.
All the program and files used or mentioned in this paper are
available at \href{http://www.cliplab.org/papers/sCASP-ICLP2020/}{\nolinkurl{http://www.cliplab.org/papers/sCASP-ICLP2020/}}.

\redbefsect
\section{Background:  s(CASP)}
\label{sec:background:-scasp}

\mclcommin{Make it clear why \texttt{a :- not b. b :- not a. c :- not c. } works}
\mclcommin{Dual program}

Answer Set Programming is a logic programming and modelling language
that evaluates normal logic programs under the stable model
semantics~\cite{gelfond88:stable_models}.
s(CASP)~\cite{scasp-iclp2018}, based on
s(ASP)~\cite{marple2017computing}, is a top-down, goal-driven ASP
system, that can evaluate Constraint ASP programs with function
symbols (\emph{functors}) and constraints \textbf{without}
grounding them either before or during execution.
Grounding is a procedure that substitutes program variables with the
possible values from their domain. For most classical ASP solvers,
grounding is a necessary pre-processing phase.  Grounding, however,
requires program variables to be restricted to take values in a finite
domain.  As a result, modeling continuous change is challenging for
ASP solvers, while it is easier for s(CASP).
%
Constraints 
improve both expressiveness and efficiency in logic programming.
As a result, s(CASP) is more expressive and faster than s(ASP).

An s(CASP) program is a set of clauses of the form: \smallskip

{\centering \code{h:-c$_1$, $\ldots$, c$_l$, b$_1$, $\dots$, b$_m$, not b$_{m+1}$, $\dots$, not b$_n$.} \par}

\smallskip

\noindent where \code{h} and \code{b$_1$, $\dots$, b$_n$} are atoms,
\code{not} corresponds to \emph{default} negation, and 
\code{c$_i$, $\ldots$, c$_l$} are constraints in some constraint system.

In s(CASP), and unlike Prolog's negation as failure, %
\code|not p(X)| is a constructive negation that can return bindings
for \code{X} on success.  Unlike in ASP, the variable \code{X} does
not need to appear in any positive literal.
%
%
Both are possible because s(CASP) resolves negated atoms %
\code{not b$_i$} against \emph{dual rules} of the
program~\cite{scasp-iclp2018,marple2017computing}.
Non-ground calls to \code{not p(X)} return the bindings
for which the call \code{p(X)} would have failed.
We summarize here the synthesis of the dual of a logic program $P$:
%
%
Clark's completion~\cite{Clark78} is first performed
and then De Morgan's laws are applied.

  \begin{enumerate}
  \item The $i$-th rule of the predicate defining $p/n$ is
    \lrul{p_i(\vec{x})}{B_i} ($i=1,\dots,k$).  $B_i$ is a conjunction
    of positive
    ($b_{i.j}$) or negative ($\neg b_{i.j}$) literals. Let $\vec{y_i}$ be the
    list of variables that occur in the body $B_i$ but do not occur in
    the head.  The first-order formula corresponding to the  rule is
    $\forall\vec{x}\ ( p_i(\vec{x}) \ \leftarrow \ \exists \vec{y_i} \
    B_i)$.


  \item With the rules for a predicate we construct its Clark's completion:

    \smallskip

  \begin{tabular}{lll}

    $\forall\vec{x}\ (\ p(\vec{x})$&$\longleftrightarrow$&$p_1(\vec{x})
    \lor \dots \lor p_k(\vec{x})\ )$\\

    $\forall\vec{x}\ (\ p_i(\vec{x})$&$\longleftrightarrow$&$\exists
    \vec{y_i} \ (b_{i.1} \land \dots \land b_{i.m} \land \lnot\ b_{i.m+1}
    \land \dots   \land \lnot\ b_{i.n})\ )$\\

  \end{tabular}

  \smallskip

\item Their semantically equivalent duals $\lnot p/n$, $\lnot p_i/n$ are:

  \smallskip

  \begin{tabular}{lll}
    $\forall\vec{x}\ (\ \lnot p(\vec{x})$&$\longleftrightarrow$&$\lnot (
    p_1(\vec{x})
    \lor \dots \lor p_k(\vec{x}))\ )$\\

    $\forall\vec{x}\ (\ \lnot p_i(\vec{x})$&$\longleftrightarrow$&$\lnot \
    \exists \vec{y}_i\ (b_{i.1} \land \dots \land b_{i.m} \land \lnot\
    b_{i.m+1} \land \dots \land \lnot\ b_{i.n})\ )$\\
  \end{tabular}

  \smallskip
  
\item Applying De Morgan's laws we obtain:

  \smallskip

  \begin{tabular}{lll}
   $\forall\vec{x} (\ \lnot p(\vec{x})$&$\longleftrightarrow$&$\lnot
    p_1(\vec{x})
    \land \dots \land \lnot p_k(\vec{x})\ )$\\

    $\forall\vec{x}\ ( \lnot p_i(\vec{x})$&$\longleftrightarrow$&$ 
    \forall \vec{y}_i\ (\lnot b_{i.1} \lor \dots \lor \lnot \ b_{i.m}
    \lor \ b_{i.m+1} \lor \dots \lor \ b_{i.n})\ )$
  \end{tabular}


  \medskip \noindent This provides a definition for
  $ \lnot p(\vec{x})$ via a clause with head $\lnot p_i(\vec{x})$
  for each original clause with head  $p_i(\vec{x})$.
  A construction (\code{C-forall}) to implement the universal
  quantifier introduced in the body of the dual program~\cite[Section
  3.4]{scasp-iclp2018} is provided by the metainterpreter.

\end{enumerate}
  
  Definitions for negated literals in $P$
  and for each of the newly negated literals
  are thus synthesized.  At the end of the chain,
  constraints are negated and handled by the constraint solver.
  After removing explicit quantifiers (as in Horn clauses),
%
%
  the dual program is then executed in a top-down fashion by a
  meta-interpreter~\cite{scasp-iclp2018,marple2017computing} which does
  not implement  SLD semantics. Instead, its main distinguishing
  points are:

\begin{description}
\item[Global constraints:]
  The s(CASP) compiler automatically generates a
  \code{global_constraint} predicate that captures all the global
  constraints written by the programmer (e.g., constructions of the
  form \code{:-p, q} to express that \code{p} and \code{q} cannot hold
  simultaneously).  This goal is always executed by the
  metainterpreter to ensure that models are consistent.

\item[Loop handling:]

  Two different cases are distinguished.  
  \begin{itemize}
  \item When a call eventually invokes itself and there is an odd number
    of intervening negations (as in, e.g., %
    \quad \mbox{\code{p:-q. $~$ q:-not r. $~$ r:-p.}}), the evaluation
    fails (and backtracks) to avoid contradictions of the form
    $p \land \lnot p$.
  \item When there is an even number of intervening negations, as in 
    \mbox{\code{p:-not q. $~$ q:-r. $~$ r:-not p.}}
   the metainterpreter generates two stable models,
    \code|{p, not q, not r}| and \code|{q, r, not p}|.
  \end{itemize}
  Additionally, the s(CASP) compiler detects statically
  rules of the form  %
  \code{r:-q, not r.} and introduces global constraints
  to ensure that  the models satisfy $\lnot q \lor r$, even
  if the literals \code{r} or \code{q} are not needed to solve the
  query. Therefore, s(CASP) will state that the program

  \vspace*{-.75em}
  
  \begin{multicols}{3}
\begin{lstlisting}[style=MySCASP]
p :- not q.
q :- not p.
r :- not r.
\end{lstlisting}
  \end{multicols}

  \vspace*{-1.75em}

  \noindent
  has no stable models, regardless of the initial query. 
  

\end{description}

\noindent
In addition to default negation, s(CASP) supports classical negation
to capture the explicit evidence that a literal is false. Classical
negation is defined with the prefix \code{'-'}, e.g., \code{-p(a)}
would mean that it is not the case that \code{p(a)} is true. When a
literal \code|p(X)| and its explicit negation \code|-p(X)| appear in a
program, s(CASP) automatically adds the constraint %
\code|:--p(X), p(X)| to ensure that they are not simultaneously true.

The execution of an s(CASP) program starts with a \emph{query} of the
form %
\code|?-c$_q$, l$_1$, $\dots$, l$_{m}$|, %
where \code{l$_i$} are (negated) literals and \code{c$_q$} is a
conjunction of constraint(s). This is followed by the evaluation of
the global constraints written by the user or introduced by the compiler.
The answers to the query are partial stable models where each one is a subset of a consistent stable
model~\cite{gelfond88:stable_models} including only the literals
necessary to support the query (see~\cite{scasp-iclp2018} for
details). Additionally, for each partial stable model s(CASP) returns
on backtracking the justification tree and the bindings for the free
variables of the query that correspond to the most general unifier
(\emph{mgu}) of a successful top-down derivation consistent with this
stable model.

s(CASP), available at
\mbox{\url{https://gitlab.software.imdea.org/ciao-lang/scasp}}, is
implemented in Ciao
Prolog~\cite{hermenegildo11:ciao-design-tplp-medium}.

\redbefsect
\section{Justification Trees of Goal-Directed 
 Evaluations}
\label{sec:answering-why-with}


We will show how s(CASP) execution trees can provide
concise justifications for the models supporting a query and how
they address several challenges~\cite{DBLP:journals/tplp/FandinnoS19}
arising in the context of XAI.
\begin{definition}[Minimal s(CASP) Justification Tree]
  \label{def:justification}
  \leftskip=10pt\parindent=0pt %
  Let $Q$ be a query to a CASP program $P$.  The s(CASP)
  \emph{Justification Tree} of a (partial) model of $P$ is the ordered
  list of literals in the 
  path of the successful goal-driven proof of $Q$, conjoined
  with that of the proof of the global constraint under the operational
  semantics of s(CASP). 
\end{definition}

Note that for every model returned by s(CASP), its \emph{Justification
  Tree} is \emph{minimal} because it only contains the literals needed
to support Q and the consistency of the global constraints in $P$, i.e.,
the subset of literals contained in a stable
model that are relevant to $Q$.
\mclcommin{Characterizing what s(CASP) generates as minimal is not
  clear to me... for example, in \code{a. a:- b. b. ?- a.}  there are
  two derivations of \code{a}: one for \code|{a}| and another one for
  \quad \code|{a,b}| \quad, of them being strictly smaller than the
  other one.  It is true that s(CASP) does not use unnecessary atoms
  in a given derivation, but when there are several comparable models,
  s(CASP) does not return the smallest one.  }

  \jahcommin{I agree that s(CASP) does not return the minimal partial
    models, but, the resulting justification trees are not comparable
    and for each partial model its justification tree is minimal.}

\noindent
Let us describe how an s(CASP) justification tree is to be understood.
We will follow the evaluation of an example query under s(CASP).

\begin{figure}
  \begin{subfigure}[b]{.48\linewidth}
\begin{lstlisting}[style=MySCASP]
opera(D) :- not home(D).
home(D) :- not opera(D).
home(monday).
\end{lstlisting}
\begin{lstlisting}[style=MySCASP, numbers=none, xleftmargin=0cm]
% QUERY:
?- opera(A).

% BINDINGS: 
A \= monday 
\end{lstlisting}
  \end{subfigure}
  \begin{subfigure}[b]{.48\linewidth}
\begin{lstlisting}[style=Plain]
% EVALUATION TREE:
opera(A │{A \= monday}) :-
    not home(A │{A \= monday}) :-
        not o_home_1(A │{A \= monday}) :-
            chs(opera(A │{A \= monday})).
        not o_home_2(A │{A \= monday}) :-
            A \= monday.
global_constraint.
$~$
\end{lstlisting}
  \end{subfigure}

  \begin{subfigure}{1.0\linewidth}
    \redbefmult
\begin{lstlisting}[style=Plain]
% MODEL: 
{ opera(A │{A \= monday}), not home(A │{A \= monday}) }

\end{lstlisting}
  \end{subfigure}
\caption{Code, query, bindings, model, and evaluation of \code|opera.pl|.}
\label{fig:opera}
\redaftfig
\end{figure}


\begin{example}
\label{exa:opera}
\leftskip=10pt\parindent=0pt %
  The program in Fig.~\ref{fig:opera}, adapted from
  \cite{DBLP:journals/eswa/GarciaCRS13}, represents Bob's plans.
  %
  Bob 
  either goes to the opera or stays home.  Bob will stay at home on
  Monday and therefore
  %
  the query \code{?-opera(A)} returns a
  partial stable model with the constraint 
  \code{A \= monday}, 
  meaning that \code{opera(A)} is valid for any \code{A}
  except \code{monday}.
  %

  Constrained/unbound variables can appear in the top-down derivation
  steps, shown on the right.
  The notation %
  \code+opera(A │{A \= monday})+ expresses that \code{opera(A)} is
  valid when \code{A $\not=$ monday}.  Following the evaluation trace,
  the query \code{?-opera(A)} holds if \code{not home(A)} does. 
  Being a negative literal, it needs to be resolved 
%
  against the dual\footnote{The dual of \code{opera.pl} is available
    at \myurl{opera_dual.pl} for the reader's convenience.} of
  \code{home/1}:

  \mclcommin{I am not sure we can add 15+ pages of appendices in
    EPTCS.  They do not have the structure of a TPLP paper, with
    separate additional material.  I would suggest to leave all the
    material in an URL and give its address in the paper.}

\redbefmult
\begin{multicols}{2}
\begin{lstlisting}[style=MySCASP, xleftmargin=.85cm]
not home(A) :- not o_home_1(A),
               not o_home_2(A).
not o_home_1(A) :- opera(A).
not o_home_2(A) :- A \= monday.
\end{lstlisting}
\end{multicols}
\redaftmult

\noindent
\code{not home(A)} succeeds because the two goals in its body succeed:
(i) \code{not o_home_1(A)} holds because s(CASP) detects that
\code{opera(A)} was called through an even number of intervening
negations.  This is marked by wrapping it with \code{chs/1}.
(ii) \code{not o_home_2(A)} succeeds because \code{A} is a free
variable and applying the constraint \code{A \= monday} succeeds.
This constraint is automatically propagated to the initial variable.
Finally, since there are no global constraints to be checked, a
partial model is generated, including the constraint \code|{A \= monday}|.


\end{example}

Dual predicates, such as \code{not o_home_1(A)}, are synthesized by
the compiler and therefore not immediately known by the user.  While
it is possible to control which predicates appear in the
representation of an evaluation tree
(Section~\ref{sec:contr-liter-that}), they are necessary to fully
understand all the details of a justification.  The s(CASP) compiler
can output the original program together with its dual to help in this aspect.

\redbefsect
\subsection{Partial Models, Constraints, and Justifications}
\label{sec:one-vs-exponentially}

Conciseness usually makes understanding easier; the goal-directed
strategy of s(CASP) contributes to concise justifications:

\begin{itemize} 
\item 
  s(CASP) computes partial stable models 
  that contain only the literals relevant to a query.
  Accordingly, s(CASP) justification trees only need to address the
  literals that appear in the partial model.
\item When there is more than one derivation tree that supports the
  query, justifications are given separately.  If only one is
  necessary, we do not have to compute all of them.
\item Additionally, constraints 
  can concisely and finitely represent infinitely many
  models.
\end{itemize}


\begin{example}[Cont. Example~\ref{exa:opera}]
\label{exa:opera_clingo}
\leftskip=10pt\parindent=0pt %
We adapted the program in Fig.~\ref{fig:opera} to evaluate
Example~\ref{exa:opera} under 
\emph{clingo}~\cite{clingo}.  A call to a domain predicate for the
seven days of the week was added to the clauses in lines 1 and 2 to
ensure that they are safe.
%
%
Thanks to the use of constraints, s(CASP) returns a single model;
however, clingo returns 64 models.  Even with the more restrictive
query \code{?-opera(tuesday)}, clingo still generates 32 models.
%
As a consequence, an explanation framework for clingo such as
\texttt{xclingo}~\cite{cabalar20:xclingo}, based on Causal Graph
Justifications~\cite{DBLP:journals/tplp/CabalarFF14},
takes 10 times longer than s(CASP) to justify the results of
\code{opera.pl} because it needs to process every single model.

\end{example}

\redbefsect
\subsection{Justifications of Negated Literals}
\label{sec:just-negat-liter}

Real-life domains
often use rules such as ``\emph{\ldots when it is not the case
  that\ldots}''.  In these cases, justifying 
why a literal is not included in a model is necessary, especially in
a non-monotonic reasoning framework such as ASP.  
In s(CASP), classically-negated literals (those with rules specifying
when they do not hold) are evaluated and justified similarly to
positive literals, with the proviso that global constraints are added
to ensure the consistency of the model.
%
%
For default-negated literals, s(CASP) uses the dual program to
constructively determine when the default negation of a literal
succeeds using the top-down evaluation mentioned in
Section~\ref{sec:background:-scasp}.


For simplicity, we will use a propositional program to illustrate the
justifications of negated predicates.  Note that thanks to the
top-down evaluation and the usage of dual programs, s(CASP) can
combine justifications for constrained, non-ground evaluations (as in
Fig.~\ref{fig:opera}) with justifications for negated literals.



\begin{figure}
  \begin{subfigure}[b]{.65\linewidth}
  \begin{multicols}{2}
  \begin{lstlisting}[style=MySCASP]
% Treatments
intraocularLens :-
    correctiveLens,
    not glasses,
    not contactLens.
laserSurgery :-
    shortSighted,
    not tightOnMoney,
    not correctiveLens.
glasses :-
    correctiveLens,
    not caresPracticality,
    not contactLens.
contactLens :-
    correctiveLens,
    not afraidToTouchEyes,
    not longSighted,
    not glasses.
% Auxiliar
correctiveLens :-
    shortSighted,
    not laserSurgery.
tightOnMoney :-
    student,
    not richParents.
caresPracticality :-
    likesSports.

% Peter, a patient
shortSighted.
student.
likesSports.
afraidToTouchEyes.

% QUERY:
?- intraocularLens.
\end{lstlisting}
  \end{multicols}
\end{subfigure}
\begin{subfigure}[b]{.33\linewidth}
\begin{lstlisting}[style=Plain]
% JUSTIFICATION_TREE:
intraocularLens :-
   correctiveLens :-
      shortSighted,
      not laserSurgery :-
        not o_laserSurgery_1 :-
           proved(shortSighted),
           tightOnMoney :-
              student,
              not richParents.
   not glasses :-
      not o_glasses_1 :-
        proved(correctiveLens),
        caresPracticality :-
           likesSports.
   not contactLens :-
      not o_contactLens_1 :-
        proved(correctiveLens),
        afraidToTouchEyes.
\end{lstlisting}
\end{subfigure}

\begin{subfigure}{1\linewidth}
\begin{lstlisting}[style=Plain]
% MODEL:
{ intraocularLens,  correctiveLens,  shortSighted,  not laserSurgery,  tightOnMoney,  student, not glasses, caresPracticality, likesSports, not contactLens, afraidToTouchEyes }
\end{lstlisting}
\end{subfigure}

\caption{Code, query, model, and justification of an ophthalmologist
  system.}
\label{fig:oph}
\redaftfig
\end{figure}

\begin{example}\label{exa:oph}
  \leftskip=10pt\parindent=0pt %
  Consider an ophthalmologist diagnosis
  system~\cite{DBLP:journals/tplp/0001T16} (Fig.~\ref{fig:oph}) that
  encodes knowledge about optical treatments.  For a given patient,
  Peter, \code{intraocularLens} are recommended.
  Fig.~\ref{fig:oph}, bottom, shows the partial model where the
  negated literals correspond to disrecommended treatments;
  these
  contribute to making the model more understandable and also increase
  trust on the system's advice, because a doctor can check which
  treatments have been discarded. 
  The s(CASP) justification tree (Fig.~\ref{fig:oph}, right, and at
  \myurl{peter.html}) justifies this recommendation.
To help understand it, we include here the dual
of \code{laserSurgery} as generated by s(CASP):\footnote{The dual of
  the ophthalmologist system is available at~\myurl{peter_dual.pl}.}
  
\begin{lstlisting}[style=MySCASP, xleftmargin=.85cm]
not o_laserSurgery_1 :- not shortSighted.
not o_laserSurgery_1 :- shortSighted, tightOnMoney.
not o_laserSurgery_1 :- shortSighted, not tightOnMoney, correctiveLens.
\end{lstlisting}

\vspace*{.5em}

Strictly speaking, clause 2 would only need to be %
\quad \code{not o_laserSurgery_1 :-tightOnMoney}.  However, including
\code{shortSighted}, which appears negated in the previous clause,
avoids unnecessary search when generating models, and therefore it
also helps 
remove redundant justifications.%
\footnote{Generating justification trees for all the possible
  derivations boils down to not producing the optimized code shown
  above.  Since this requires additional search and execution time, it
  ought to be a command-line option.
}
Clause 3 belongs to a similar case.

  Note that although there are two ways of justifying
  \code{not laserSurgery} (i.e., using \code{tightOnMoney} or
  \code{correctiveLens}), only one is necessary and
  generated.
  \mclcomm{That is in some sense an issue, because providing other
    justifications is not possible even on backtracking.}
  %
  As mentioned before, the wrapper \code{proved/1} marks literals
  already proven whose justification sub-tree can be omitted.

%
%
   %
  %


\end{example}

Some proposals such as LABAS~\cite{DBLP:journals/tplp/0001T16} justify
all the negated literals, which may provide too much detail. Others,
such as Causal Graph
Justifications~\cite{DBLP:journals/tplp/CabalarFF14}, treat negated
literals as assumptions and give no justification for them.
s(CASP) provides a balance as it 
only justifies the negated literals needed to support the answer to
the query and avoids generating redundant justifications thanks to the
way dual rules are generated.
Note that reducing the number of unnecessary justifications is
relevant in practice:
justifying the literals that do not support the query may increase the
number of justifications (and the execution time) exponentially
w.r.t.\ size of the program~\cite{DBLP:journals/tplp/FandinnoS19}.

Additionally, the s(CASP) implementation includes ways to control
which literals should appear in the justification tree
(Section~\ref{sec:contr-liter-that}).  This makes it possible to
tailor the contents and size of the justification tree to better
adapt it to specific cases.

\redbefsect
\subsection{Justifications of Global Constraints}
\label{sec:just-glob-constr}

%
Global constraints are introduced by the s(CASP) compiler
(Section~\ref{sec:background:-scasp}) to ensure model consistency in
the presence of classically-negated literals and 
of rules such as \quad \code{p:-q, not p.}\quad  Consistency rules written by
the user are also added to the global constraints.
Therefore, it is necessary to generate justifications to explain how
models are compliant with global constraints.

\begin{figure}
  \begin{multicols}{2}
\begin{lstlisting}[style=Plain]
% JUSTIFICATION_TREE:
opera(A │{A \= monday,A \= tuesday}) :-
  not home(A │{A \= monday,A \= tuesday}) :-
    not o_home_1(A │{A \= monday,A \= tuesday}) :-
      chs(opera(A │{A \= monday,A \= tuesday})).
    not o_home_2(A │{A \= monday,A \= tuesday}) :-
      A \= monday.
global_constraint :-
  not o_chk :-
    not o_chk_1 :-

      forall(C,not o_chk_1(C)) :-
        not o_chk_1(B │{B \= tuesday}) :-
          not baby(B │{B \= tuesday}) :-
            not o_baby_1(B │{B \= tuesday}) :-
              B \= tuesday.
        not o_chk_1(tuesday) :-
          baby(tuesday),
          not opera(tuesday) :-
            not o_opera_1(tuesday) :-
              home(tuesday) :-
                chs(not opera(tuesday)).
\end{lstlisting}
  \end{multicols}
    \caption{s(CASP) justification of \code|opera.pl| with a global
      constraint.}
    \label{fig:opera_const}
\redaftfig
\end{figure}

\vspace*{1em}

\begin{example}[Cont. Example~\ref{exa:opera}]
  \label{exa:opera_const}
  \leftskip=10pt\parindent=0pt %
  Let us modify the code in Fig.~\ref{fig:opera} 
  by adding:
  

\begin{lstlisting}[style=MySCASP, xleftmargin=.85cm]
:- baby(D), opera(D).   % When Bob's best friend comes with her baby, it is
                        % not a good idea to take the baby to the opera.
 baby(tuesday).         % They come on Tuesday.
\end{lstlisting}

\vspace*{.5em}

The answer for \code{?-opera(A)} is now %
\code{A \= monday, A \= tuesday} because Bob always stays home on
Monday and Bob's best friend comes with her baby on Tuesday.
Fig.~\ref{fig:opera_const} shows the s(CASP) justification tree
including the global constraint checks. The first part of the
justification tree is similar to that in Fig.~\ref{fig:opera}, left,
with the addition that variable \code{A} is constrained not to be
equal to \code{tuesday}.
To follow the justification of the user constraints, let us 
remember that \quad \code{:-baby(D), opera(D).}\quad is
$\forall x \cdot \neg(baby(x) \land opera(x))$.  This is compiled into

  \begin{minipage}{.85\linewidth}
    \begin{multicols}{2}
\begin{lstlisting}[style=MySCASP,xleftmargin=.55cm]
not o_chk_1 :- 
   forall(D, not o_chk_1(D)).
not o_chk_1(D) :- not baby(D).
not o_chk_1(D) :- baby(D), not opera(D).
\end{lstlisting}
    \end{multicols}
  \end{minipage}

  \vspace*{.5em}

  \noindent
  where the compiler introduced the predicate
  \code{forall/2}~\cite[Section
  3.4]{scasp-iclp2018}
  that succeeds if for all possible values of \code{D} present in the
  program, the predicate \code{not o_chk_1(D)} holds.
  The justification tree shows that:
  \begin{itemize}[nosep, leftmargin=3em]
  \item \code{not o_chk_1(B)} succeeds for \code{B \= tuesday} because
    \code{not baby} succeeds for \code{B \= tuesday}. 
  \item \code{not o_chk_1(tuesday)} holds because \code{baby(tuesday)}
    is true and \code{not opera(tuesday)} succeeds (the variable
    \code{A} of \code{opera/1} can be restricted to be different from
    \code{tuesday}). The resulting constraint is %
    \code|{A \= monday, A \= tuesday}|.
  \end{itemize}
\end{example}

 \redbefsect
\subsection{Controlling the Literals that Appear in an s(CASP)
  Justification Tree}
\label{sec:contr-liter-that}

%


ASP systems (and also s(CASP)) provide mechanisms to control which
literals should be shown in a model.
Controlling which literals should appear in a 
justification tree is more challenging.  A decision based on hierarchy
and predicate dependencies is not satisfactory, as we observed that in
some cases we may want to hide some literals but not their children in
order to inspect the support for some conclusions.  This means taking
special care for some situations (e.g., having to remove repeated
siblings) while providing an interface to facilitate a
flexible control.




The current implementation supports three levels of detail
(\code{short, mid, long}) plus an additional option
(\code{neg}) that makes the interpreter include the default-negated
literals:

\begin{description}[noitemsep, wide=.75cm, leftmargin=1.5cm]
\item[\texttt{short}] shows the (negated) literals selected with
 \texttt{\#show} directives.
\item[\texttt{mid}] adds 
  the rest of user-defined predicates (positive and/or classically
  negated).
\item[\texttt{neg}] (only with \code{short} or \code{mid})
  shows also the default-negated versions of the displayed predicates.
\item[\texttt{long}] generates the complete s(CASP)
  justification tree, including auxiliary predicates, \code{forall},
  and built-ins.
\end{description}

\noindent
s(CASP) can generate justification trees as plain text or as
expandable \texttt{html}, accessible through links in this paper.
In each of these cases, the three options mentioned above are
available.



%
%
%

\redbefsect
\section{Presenting s(CASP) Justification Trees using Natural Language}
\label{sec:using-natur-lang}

Justification trees lend themselves to the generation of natural
language explanations quite directly.  This is useful for domain
experts that may not be familiar with logic rules.
Predefined generic patterns are used to generate simple translations.
Additionally, more involved explanations, tailored to each particular case, can
be produced by adding natural-language explanations to the program.
These offer a human-readable meaning of the program at the individual
rule level and can control which literals have to be explained.  They
can be applied both to plain text and HTML generation.


\begin{figure}
\begin{subfigure}{1\linewidth}
\begin{lstlisting}[style=Plain]
'opera' holds (for A), with A not equal monday, because
    there is no evidence that 'home' holds (for A), with A not equal monday, because
        'rule 1' holds (for A), because
            it is assumed that 'opera' holds (for A), with A not equal monday.
        'rule 2' holds (for A), because
            A is not equal monday.
\end{lstlisting}

  \redbefmult
  \caption{Justification in English using predefined patterns.}
  \label{fig:opera_nl}
\end{subfigure}
\begin{subfigure}{1\linewidth}
\begin{lstlisting}[style=Plain]
Bob goes to the opera on a day A not equal monday, because
    Bob does not stay at home on A not equal monday, because
        'rule 1' holds (for A), because
            it is assumed that Bob goes to the opera on a day A not equal monday.
        'rule 2' holds (for A), because
            A is not equal monday.
\end{lstlisting}

  \redbefmult
  \caption{Justification in English using user messages.}
  \label{fig:ope_nl}
\end{subfigure}
  \caption{s(CASP) justification tree for \code|opera.pl| in natural language.}
  \redaftfig
\end{figure}

\redbefsect
\subsection{Predefined Natural Language Patterns}


%
Fig.~\ref{fig:opera_nl} shows the s(CASP) pseudo-natural language
justification for the program and query in Fig.~\ref{fig:opera} using
only predefined patterns.  Each line corresponds to a literal in the
justification tree shown before and is generated as follows:

\begin{description}
\item[\texttt{name(Arg1, ..., Argn)}:]
  \fcolorbox{blue}{white}{\emph{\code|'name'| holds (for
      arg$_1$,$\dots$, and arg$_n$)}} where \emph{arg$_i$} are either
  the run-time value for argument $i^{th}$ or a variable name. For
  constrained variables, the constraints are also translated and
  shown.

\item[\texttt{not name/n}:] \fcolorbox{blue}{white}{\emph{there is no evidence that}}, followed
  by the pattern for \code{name/n}.
  
\item[\texttt{-name/n}:] \fcolorbox{blue}{white}{\emph{it is not the
      case that}}, followed by the pattern for \code{name/n}.
\end{description}


\noindent
The neck \code{':-'} is translated into \emph{`because'} and the comma
\code{','} into \emph{`and'}. 
The mark \code{chs(name/n)} (a loop with an even number of negations
has succeeded) is translated into \emph{`it is assumed that'}
followed by the pattern for \code{name/n}, and the mark
\code{proved(name/n)} (the literal has already been proved) is
translated by adding \emph{`already justified'} to the pattern for
\code{name/n}.
For brevity, we will skip the description of the translation patterns
for \code{forall}, the built-ins, and the auxiliary predicates of the
dual rules.  

It is interesting to remark that, since the evaluation tree chains the
application of program rules, these predefined patterns can also be
applied to translate the program code into natural language --- the
only difference is that the neck \code{':-'} is translated into
\emph{`if'}.  The translation of the dual program of \code{opera.pl}
from Example~\ref{exa:opera} (including the dual rules and the global
constraints) is available at \myurl{opera_dual_NL.pl} for
the reader's convenience.

\redbefsect
\subsection{User-Defined Natural Language Patterns}
\label{sec:scasp-just-natur}
Customizing the predefined patterns makes it possible to better
describe the meaning of the predicates.
%
Our framework provides this possibility
%
%
by allowing structured comments to be added to the literals (either positive or
negative) of the programs.\footnote{There other proposals, similar in
  spirit, that allow adding information to programs without changing
  their meaning, e.g. those of Ciao
  Prolog~\cite{hermenegildo11:ciao-design-tplp-medium} or
  Clingo~\cite{clingo}.} Let us consider the two following examples:


\begin{lstlisting}[style=Plain]
#pred opera(D) :: 'Bob goes to the opera on @(D:day)'.
#pred not home(D) :: 'Bob does not stay at home on @(D)'
\end{lstlisting}

\vspace*{.5em}

\noindent
Each explanation of a literal is introduced with the directive
\code{#pred} followed by the (negated) literal and a 
pattern indicating how to translate it.
%
%
The marks \code{@(D)} indicates where the values of the
arguments of the literals should appear.  A qualification such as
%
\code{@(D:day)} gives additional information on the meaning of the
variable that is used to generate a more informative message depending
on the instantiation state of the
variable. Table~\ref{tab:annotations} shows an intuitive explanation.


\begin{table}
\begin{center}
\caption{Translation results with and without user annotations.}
\label{tab:annotations}
  \begin{tabular}{lclcl}
\toprule
 \textbf{Var.\ State}&\textbf{Mark}&\textbf{Translation} &\textbf{Mark}&\textbf{Translation} 
\\\midrule
  Free & \texttt{@(D)} & \emph{D} & \texttt{@(D:day)} & \emph{D, a day} \\
  Constrained & \texttt{@(D)} & \emph{D not equal monday} & \texttt{@(D:day)} & \emph{a day D not equal monday } \\
  Ground & \texttt{@(D)} & \emph{monday} & \texttt{@(D:day)} & \emph{the day monday } 
\\\toprule
\end{tabular}
\end{center}
\redafttab
\end{table}


Messages can be defined for different instantiation patterns of the
same literal, as in 

\begin{lstlisting}[style=Plain]
#pred is(pregnancy) :: 'the patient is pregnant or planning to get pregnant'.
#pred is(stage(S)) :: 'the patient is in ACCF stage @(S)'.
#pred is(E) :: 'the patient is @(E)'.
\end{lstlisting}

\vspace*{.5em}

Literals are scanned from top to bottom and the message associated with the
first matching head is used.  That makes it possible, for example, to
write different explanations for every clause of a predicate (from
more particular to more general), as it is common (see the case above)
that every clause treats a different case.
%
Translations for default-negated and classically-negated literals can
also be provided.

Fig.~\ref{fig:ope_nl} shows the s(CASP) justification tree for
\code{opera.pl} (Fig.~\ref{fig:opera}) using user-defined explanations.
The complete s(CASP) justification tree and the dual program,
translated following the user-defined patterns, are available at
\myurl{opera_just_userNL.pl} and \myurl{opera_dual_userNL.pl}.

s(CASP) generates
natural language explanations with the command-line option
\code{--human}.
The \code{short} version also shows the (negated) literals with
user-defined predicates, even if they are not selected in the
\code{#show} directives.

\redbefsect
\section{A Real Application: the Physician Advisor System}
\label{sec:real-appl-phys}

We have evaluated the s(CASP) justification generation
with the Physician Advisor System~\cite{chen2016physician} (PAS), an
ASP program that recommends treatment choices for chronic heart
failure (CHF).  The system encodes near 80 pages of rules
into an ASP program.
Medical advisory systems are relevant applications in the context of
XAI, and the management of chronic diseases, such as CHF
and others, is a major problem in health care.
%
%
The PAS has a modular implementation:

\begin{itemize}[nosep]
\item The main module of the PAS (available
  at~\myurl{PAS_rules.pl}) implements the knowledge patterns
  described in~\cite{chen2016physician}.
\item The guide module (available at~\myurl{PAS_guide.pl})
  implements the CHF Guide.
\item The patient module (available at~\myurl{PAS_profile.pl})
  contains a patient profile: demographic data, assessment evidences,
  particular contraindications, previous diagnosis, illness,
  medication history, and different measurements.
\end{itemize}

\begin{figure}[tb]
\begin{lstlisting}[style=Plain, basewidth=.45em]
the treatment ace_inhibitors has been chosen, because
    it is a recommendation to use ace_inhibitors, because
        the patient is in ACCF stage C, and
        the patient is diagnosed with heart failure with reduced ejection fraction, because
            there is a measurement of lvef of 0.35.
        there is no evidence that there is a danger in taking ace_inhibitors, because
            there is no evidence that the patient has a history of angioedema, and
            there is no evidence that the patient is pregnant or planning to get pregnant.
    there is no evidence that the treatment ace_inhibitors is excluded, because
        the treatment diuretics is concomitant if ace_inhibitors is chosen, and
        it is a recommendation to use diuretics, because
            the patient is in ACCF stage C, justified above, and
            the patient is diagnosed with heart failure with reduced ejection fraction, justified above, and
            there is no evidence that there is a danger in taking diuretics.
        the treatment diuretics has been chosen, because
            it is a recommendation to use diuretics, justified above, and
            there is no evidence that the treatment diuretics is excluded, and
            there is no evidence that diuretics is discarded.
        the treatment aldosterone_antagonist is incompatible with ace_inhibitors, and
        aldosterone_antagonist is discarded, and
        the treatment arbs is incompatible with ace_inhibitors, and
        arbs is discarded.
    there is no evidence that ace_inhibitors is discarded.
The global constraints hold, because
    the global constraint number 1 holds.
\end{lstlisting}
    \caption{s(CASP) justification tree of the Physician Advisor System.}
  \redaftfig
    \label{fig:ace}
\end{figure}

\noindent
To provide readable recommendations,
we extended PAS with explanations for the relevant predicates
(available at~\myurl{PAS_rules.pred.pl}, \myurl{PAS_guide.pred.pl},
and \myurl{PAS_patient.pred.pl}).
%
With them, s(CASP) can generate concise justification trees in natural
language.

\redbefsect
\subsection{Expressiveness of Justifications}

Let us study one example: for the query \code{?-chose(ace_inhibitor)}
there is a unique minimal stable model. Fig.~\ref{fig:ace} shows the
natural language version of the s(CASP) justification tree for
that query.\footnote{Other relevant queries to the PAS are listed
in~\myurl{PAS_query.pl} and at
\myurl{ace_inhibitors.html}, \myurl{beta_blockers.html}, and
\myurl{anticoagulation.html}.}
  The following points are worth mentioning:

\begin{itemize}

\item The justification tree supports the choice of
  \code{ace_inhibitor} based on evidence (positive literals), e.g.,
  ``the patient is in ACCF stage C'', and on the absence of
  counter-evidence (negated literals), e.g., ``there is no evidence
  that there is a danger in taking ace\_inhibitors''.
  As mentioned before, providing justifications of negated literals is
  important because a doctor may want to double-check them.



\item The justification includes only the recommendations and choices
  of treatments that are required to support the choice of
  \code{ace_inhibitor}, e.g., the treatment with \code{diuretics} is
  chosen because it is concomitant when \code{ace_inhibitor} is
  chosen.

\item The justification also explains that some treatments have been
  discarded due to incompatibilities.
  This information is essential since it would avoid future errors by
  preventing the use of \code{aldosterone_antagonist} or \code{arbs}.
  \mclcommin{Where do we \emph{see} the
    incompatibility?}\jahcommin{Should we include the partial model in
    the text?}\mclcommin{I am not sure that this will help.  Again, we
    are referring to the justification tree.  Where do we see that
    incompatible treatments were discarded?}\jahcommin{It can be see
    also in the justification tree.}



\item s(CASP) supports dense domains which can directly represent
  physical quantities, such as the measurements of \emph{lvef}
  (\code{0.35} in this example).  This makes it possible to directly
  encode existing regulations and use these measurements in the
  justifications.



\end{itemize}

\medskip

\begin{wrapfigure}{r}{0.5\linewidth}
  \vspace{-1.5em}
\begin{lstlisting}[style=Plain]
    >> chose(ace_inhibitors)	
      |__recommendation(ace_inhibitors)
         |__reason(ace_inhibitors)
            |__evidence(accf_stage_c)
            |__diagnosis(hf_with_reduced_ef)
               |__measurement(lvef,35)
\end{lstlisting}
\caption{Fragment of \texttt{xclingo} justification.}
\label{fig:xclingo-tree}
\end{wrapfigure}
We compared s(CASP) with
\texttt{xclingo}~\cite{cabalar20:xclingo}.  \texttt{xclingo} does
not generate justifications for negated literals.  The justification
of the literal \code{chose(ace_inhibitors)} (generated using the
\code{
\code{discarded(T)}) is shown in Fig.~\ref{fig:xclingo-tree}.

It can be noted that the information related to negated literals, such
as \code{not contraindication/1} and %
\code{not exclude/1}, is absent.  As a consequence, the justification
generated by \texttt{xclingo} does not fully explain why
\code{diuretics} is chosen or why \code{aldosterone_antagonist} and
\code{arbs} are discarded in every model.\footnote{The
  \texttt{xclingo} justification for one of the models for the query
  \code{?-chose(ace_inhibitors)}, including the justification of every
  literal \code|chose/1| and \code|discarded/1| in the answer set, is
  available at~\myurl{PAS_xclingo_just.pl}.} Since clingo does not
support decimals, the values of \emph{lvef} have been normalized.

\redbefsect
\subsection{Performance Evaluation}



We have evaluated the additional cost of generating justifications in
s(CASP) and also compared the performance of our approach with that of
other similar systems using, again, the PAS system.  We used a Debian
4.19 machine with a Xeon E5640 at 2.6GHz.


  
\begin{table}[tb]
  \centering
  \caption{Performance comparison of w.r.t.\ evaluation
    w.o. justification tree.}
  \label{tab:eval}
 
  \begin{tabular}{p{5cm}rrrr}
    \toprule
    & \multicolumn{2}{c}{} & \multicolumn{2}{c}{\code|--html|} \\\cline{4-5}
                   &\code|--plain|&\code|--human|&\code|--plain|&\code|--human|\\
    \midrule
    \code|--tree --short           |   & 1.06   & 1.06   & 1.12   & 1.12    \\
    \code|--tree --short --neg     |   & 1.06   & 1.07   & 1.12   & 1.13    \\
    \code|--tree --mid             |   & 1.06   & 1.06   & 1.12   & 1.12    \\
    \code|--tree --mid --neg       |   & 1.06   & 1.07   & 1.12   & 1.13    \\
    \code|--tree --long            |   & 1.07   & 1.08   & 1.14   & 1.15    \\
    \bottomrule
  \end{tabular}
\redafttab
\end{table}


We used the query \code{?-chose(ace_inhibitors)} to test the
performance of the generation of justifications.  Generating
justifications in s(CASP) has a very small impact on execution time
which we show in Table~\ref{tab:eval} as speed-down (the larger the
number, the slower the execution) and which is, at most, a 15\% slower,
which can be considered perfectly acceptable.  This is also true for
absolute speed, since the execution without justifications takes 340
ms.


  %

For the same query,
the program has 1024 models, because treatments unrelated with the
query may be selected or not.  This makes \texttt{xclingo} need 1'53''
to justify
selected and discarded treatments, while s(CASP) uses only 0.35
seconds, as only one partial model has to be
generated.

\redbefsect
\section{Conclusion}
\label{sec:concl-future-work}

We showed how s(CASP) can generate justifications for Constraint
Answer Set programs, preventing generating excessively many
justifications thanks to its ground-free, top-down evaluation strategy
and the use of constraints. It can also justify negated literals and
global constraints.

To the best of the authors' knowledge, this approach is the only one
that can provide full justifications in natural language for  ASP programs,
including constraints and negated literals.
%
Our approach also makes it possible to control which literals should
appear in the justification tree, improving the readability of the
justifications. %


As part of our future work, we want to explore how s(CASP)
justifications can be used to bring XAI principles to a class of
knowledge representation and reasoning systems, namely those based on
non-monotonic logic with a stable model semantics.  These can range
from rule-based systems capturing expert
knowledge~\cite{chen2016physician} to ILP systems that generate ASP
programs~\cite{DBLP:conf/aaai/ShakerinG19-short} or concurrent
imperative programs based on behavioral, observable
specifications~\cite{DBLP:conf/lopstr/VaranasiSMG19}.

\redbefsect
\bibliographystyle{eptcs}

\label{lastpage}

\end{document}